\documentclass{aa}  
\usepackage{graphicx}
\usepackage{txfonts}
\usepackage{makecell}

\begin{document} 

\title{Searching for electromagnetic counterpart of LIGO gravitational waves in the \textit{Fermi} GBM data with ADWO}

   
   \author{Zsolt Bagoly\inst{1}
    	  \and
		  Dorottya Sz\'ecsi\inst{2} 
          \and
          Lajos G. Bal\'azs\inst{1,3}
          \and
          Istv\'an Csabai\inst{1}
          \and
          Istv\'an Horv\'ath\inst{4}
          \and
          L\'aszl\'o Dobos\inst{1}
          \and 
          J\'anos Lichtenberger\inst{1,5}
          \and
   		  L. Viktor T\'oth\inst{1}            
          }
   \institute{E\"otv\"os University, Budapest, Hungary.\\
              \email{zsolt.bagoly@elte.hu}
        	\and 
             Argelander-Institut f\"ur Astronomie der Universität Bonn, Germany
         	\and
             Konkoly Observatory, RCAES, Hungarian Academy of Sciences, 
          	\and   
             National University of Public Service, Hungary 
          	\and
          	 Geodetic and Geophysical Institute, RCAES, Hungarian Academy of Sciences
             }
   \date{Received March 18, 2016; accepted xxxx, 2016}

  \abstract
  {}
   {The \textit{Fermi} collaboration identified a possible electromagnetic counterpart of the gravitational wave event of September 14, 2015. Our goal is to provide an unsupervised data analysis algorithm to identify similar events in \textit{Fermi}'s Gamma-ray Burst Monitor CTTE data stream.} 
   {We are looking for signals that are typically weak. Therefore, they can only be found by a careful analysis of count rates of all detectors and energy channels simultaneously. Our Automatized Detector Weight Optimization (ADWO) method consists of a search for the signal, and a test of its significance.}
   {We developed ADWO, a virtual detector analysis tool for multi-channel multi-detector signals, and performed successful searches for short transients in the data-streams. We have identified GRB150522B, as well as possible electromagnetic candidates of the transients GW150914 and LVT151012.} 
   {ADWO is an independently developed, unsupervised data analysis tool that only relies on the raw data of the \textit{Fermi} satellite. It can therefore provide a strong, independent test to any electromagnetic signal accompanying future gravitational wave observations.}

   \keywords{gamma rays: general -- gravitational waves -- (stars:) gamma rays bursts: general  -- (stars:) gamma rays bursts: individual }

   \maketitle
\section{Introduction}

We present a new method to search for non-triggered, short-duration transients in the data-set of the \textit{Fermi} Gamma-ray Space Telescope's Gamma-ray Burst Monitor (GBM) \citep{2007JPhCS..60..115C, 2009ApJ...702..791M}. The method, called Automatized Detector Weight Optimization (ADWO), combines the data of all available detectors and energy channels, identifying those with the strongest signal. This way, we are able to separate potential events from the background noise and present the statistical probability of a false alarm. Although it is possible to apply our ADWO method to look for non-triggered short gamma-ray bursts (SGRBs), ADWO works the best if a potential event at a given time (and, if available, a given celestial position) is provided as an input. Thus, ADWO is ideal to search for electromagnetic (EM) counterparts of gravitational wave (GW) events, when the time of the event is well known from the GW-detectors' observation. 

	On September 14, 2015 at 09:50:45.391 UTC the two detectors of the advanced Laser Interferometer Gravitational-Wave Observatory (LIGO) simultaneously observed a
transient gravitational-wave signal GW150914 \citep{2016PhRvL.116f1102A}. The signal is originated from the merger of a binary black hole (BBH) system at low
redshift ($z\simeq 0.1$)
\citep{2016ApJ...818L..22A}.

	GBM observations 
revealed a weak transient source above 50 keV, 0.4 s after the GW event, with a false alarm probability of 0.0022 \citep{2016arXiv160203920C}.  This
weak transient, with a duration of $\approx1$~s, does not appear to be connected with any other
previously known astrophysical, solar, terrestrial, or magnetospheric activity.
Its localization is ill-constrained but consistent with the direction of
GW150914.  The duration and spectrum of the \textit{Fermi} transient event suggest that the radiation was arriving at a large angle relative
to the direction where the Fermi Large Area Telescope (LAT) was pointing. 

The electromagnetic transient was a result of a custom pipeline looking for prompt gamma-ray counterparts in GBM \citep{2015ApJS..217....8B,2013PhRvD..87l3004K}, optimized for LIGO/Virgo GW candidate events. The automatic GBM pipelines (looking for GRBs) did not find any transients.

	Neither the \textit{Fermi} LAT observation \citep{2016arXiv160204488F} above 100 MeV nor the
partial Swift follow-up \citep{2016MNRAS.tmpL..45E} in the X-ray, optical and UV bands, nor the INTEGRAL observations \citep{2016ApJ...820L..36S} in the  gamma-ray and hard X-ray bands found any potential
counterparts to GW150914, they only provide limits on the transient counterpart activity.
  
However, from a theoretical point of view, EM counterparts such as short duration GRBs associated with GW events are not excluded. 
Recently, \cite{2016ApJ...821L..18P} proposed a scenario where a double black hole merger is accompanied by a SGRB. The evolution of the system starts with two low-metallicity massive
stars that are orbiting around each other \citep{2009A&A...497..243D,2016A&A...588A..50M,2016MNRAS.458.2634M}. Their orbit is so tight initially that their
rotational periods are synchronized with the orbital period. Due to
the fast rotation, these stars evolve homogeneously and never expand 
\citep[as described by][for single, homogeneously evolving stars]{2015A&A...581A..15S}. This way, the stars avoid
the supergiant phase and thus a common envelope evolution, which reduces the theoretical uncertainties involved. Assuming that 
(at least) one
of the supernova explosions leaves a long-lived disk behind, \cite{2016ApJ...821L..18P} predict that this scenario leads to a relativistic jet to be launched during the merger of the black holes.  
The burst-duration timescale they derive from
their models is in the order of 5~ms.
In light of these theoretical models that predict not only the
existence of black hole mergers but even the subsequent production of a SGRB, it is quite reasonable to look for EM transients of any possible gravitational wave detection.

LVT151012, the second GW candidate transient event occurred on
October 12, 2015 at 09:54:43.555 UTC \citep{2016arXiv160203839T, 2016arXiv160604856T}.  They report a false alarm probability of 0.02, and
consider it not to be low enough to confidently claim this event as a real GW signal.

This paper is organized as follows:
in Section 2 we describe our method,
in Section 3 we test our ADWO method with the short-duration GRB150522B and in
Section 4 with the SGRB-like signal that accompanied the GW150914 event. 
We find that our analysis of these signals are in accordance with the results of \citet{2016arXiv160203920C}. In Section 5 we apply ADWO to look for a potential EM counterpart of the event LVT151012.

\section{Input data and methods}

\subsection{\textit{Fermi} GBM overview}

	The \textit{Fermi} GBM includes two sets of detectors: 12 thallium activated Sodium Iodide (NaI(Tl)) and two Bismuth Germanate (BGO) scintillation detectors
\citep{2009ApJ...702..791M}.  
	The NaI(Tl) detectors measure the low-energy spectrum (8~keV to $\sim 1$~MeV) while the BGO detectors have an energy range of $\sim 200$~keV to $\sim 40$~MeV. 
The effective area of the detectors varies with the photon energy and the angle of incidence, with a maximum of
$\sim 100$~cm$^2$ (NaI(Tl)) and $\sim 120$~cm$^2$ (BGO).

	Signals from the photomultipliers are analyzed on-board, and the pulse height analysis (PHA) converts the peak heights into 128 PHA channels. The signal distribution in these PHA channels as a
function of the incoming photon energy and geometry is described by the detector response matrix (DRM).  
The DRMs contain the
effective detection area as the function of the angular dependence of the
efficiency, energy deposition and dispersion, detector non-linearity, as well as the atmospheric and spacecraft scattering. 
The PHA distribution is usually wide for high-energy photons (especially above
$\sim~1$~MeV), as some photons will scatter prior to detection.  The DRMs
are provided as a standard data product for each GBM trigger, but neither the program nor 
the data are public.

	It is important to note that the 128 PHA channels have different
energy ranges from detector to detector, according to the detector's actual setup.
The PHA channels are aggregated into different data products, e.g. CTIME data,
which consist of accumulated spectra from each detector with a 8-channel energy
and 64/265~ms time resolution.

	A GBM trigger occurs when the count rates of two or more detectors exceed the
background with a given threshold ($4.5-7.5\sigma$). The trigger algorithms
include four energy ranges ($25-50$~keV, $50-300$~keV, $100-300$~keV, and
$>~300$~keV) and ten timescales (from 16~ms to 8.192~s).  A total of 120
different trigger algorithms can be specified, from which usually $\sim$~75 
operate simultaneously.

\subsection{Automatized Detector Weight Optimization (ADWO)}

	The basic problem of the event analysis is to find the parameters of an event
in multi-detector multi-channel time series when the approximate time and
direction of the expected signal are given. 
To calculate the significance of such an event as described by PHA counts, one
should  take the typical background noise and the spectral model into
account. To obtain the background-induced PHA counts, the assumed synthetic spectrum 
is multiplied by the DRM and binned.
This is then compared to the PHA counts derived from the combination of the
signal and the background with tools such as XSPEC for fitting 
Gaussian signals using $\chi^2$, and C-Stat for Poisson signals \citep{1996ASPC..101...17A}.

	Contrary to this detection method, here we do not assume the
event time, only a possible time interval is given. Our goal is to find the
strongest weights and the best time position in this interval using a
weighted signal from the multi-detector multi-channel continuous data.  
The simplest method would be to compare the sum of the count
rates within and outside the given time interval. This approach, however, is not the most effective one in a multi-channel multi-detector environment, since for a maximum signal-to-noise ratio
usually only those detectors should be summed (selected for the analysis) which
produce the strongest signals. Noisy energy channels and not illuminated detectors
with very low DRM should either not be taken into account, or only with a low weight. A further complication arises from the fact that we know neither the direction of the event (and, therefore, if a given detector is illuminated or not), nor the spectra.

	Our solution for these problems is the following: we give different weights to different
energy channels ($e_i$) and detectors ($d_j$), and optimize the Signal's Peak to Background's Peak Ratio (SPBPR).
The weights are positive and
normalized as $\sum e_i=1, \sum d_j=1$.  We do not restrict these weights any
further, i.e. we do not include any information about the DRM (which we do not know anyway, without
any spectral and directional information).

	If the background subtracted intensity in the
$j$th detector's $i$th energy channel is $C_{ij}(t)$, we define our composite signal as
$S(t)=\sum_{i,j} e_i d_j C_{ij}(t)$.  The Signal's Peak is the maximum of $S(t)$
within the given time interval, and the Background's Peak is the maximum of $S(t)$
outside this interval. Our goal is to maximize the ratio of these two maximums. The best weights 
will be built up by iteration, maximizing SPBPR as a function of $e_i$ and $d_j$. These $e_i$ and $d_j$ weights create an optimal filter among
the spectra and detectors. 
The algorithm will provide not only maximum value of SPBPR, but will search the best weights and the exact time of this maximum, within the pre-defined interval.

We call this algorithm the Automatized Detector Weight Optimization.  ADWO  is similar to the GRB satellites' triggering mechanism, but includes several improvements. 
For example, while \textit{Fermi}'s trigger algorithm selects the $e_i$ and $d_j$ factors
to be $0$ or $1$, here we allow intermediate values too.  Additionally, the condition that at least two detectors exceed a threshold simultaneously, is not required anymore, since the ADWO algorithm
will automatically produce the best $d_j$ weights. For a signal with time-evolving spectrum, ADWO will determine the best trigger time position.

We applied Matlab's/Octave's {\em fminsearch} routine to find the maximum via the Nelder \& Mead Simplex algorithm.  The algorithm always started from an equal weights position.  The analysis of $\approx 100$ signal data points against the $\approx 10^4$ background data points with 80 dimensional data took several minutes on a 4-core Intel i7 processor with 8GB memory, depending on the linear algebra packages used by the programs. After the search converged, the differences of the weights on the final simplex were below $10^{-4}$ (the sum of the weights is 1).
The sample Matlab/Octave code is available on GitHub ({\tt https://github.com/zbagoly/ADWO}).

\subsection{Analysis of the \textit{Fermi} GBM data}

	Since November 2012, the \textit{Fermi} GBM continuous time-tagged event (CTTE)
data is present for each detector with a time precision of $2~\mu$s, in all the
128 PHA energy channels \citep{2009ApJ...702..791M}.  Here we use the
same CTIME energy channels of \cite{2016arXiv160203920C}, with limits of
4.4,~12,~27,~50,~100,~290,~540,~980 and 2000~keV (denoted with $e_1 \dots e_8$, resp.).
Since we look for spectrally hard events, we use only the
upper 6 energy channels in the 27-2000~keV range ($e_3 \dots e_8$).  The
exclusion of the low energy channels also reduces the background contamination
from soft particle events, such as Cygnus X-1 and other weak variable X-ray sources, since
their flux is usually small above 27~keV. 

	All the 12 NaI(Tl) and both BGO detectors were included in the analysis.  Since the BGO
detectors' low energy PHA channels start above $100$~keV, the
corresponding 27-100~keV energy channels are empty. Overall, we have
$6\times14-2\times 2=80$~non-zero time series.

	For each detector and for each channel, the CTTE
$2~\mu$s event data is filtered with a 64~ms wide moving average filter at 1~ms steps, 
producing the $C_{ij}(t)$ light curve.  
This filtering is important as
the photon event data are quite sparse (the intensity is quite low; for the GW150914 event there is, on average $\approx 5.8$~ms between photons in a given
detector and energy channel).  Our 64~ms window contains $11.2$~photons on average, while this window size corresponds to the typical triggered CTIME light curve resolution.

Without filtering, the
photon-photon correlation in time that we search for would disappear.
Very narrow filters are worthless because of the sparsity constraint,
while much wider filters will smooth and filter out short transients,
lowering ADWO's sensitivity.  As a byproduct, the smoothing also acts as a
low-pass filter which reduces the Poisson noise. 
The moving average filter is the simplest choice here: e.g. using some prior knowledge about the signal's shape, a matching filter tuned to the signal would improve the sensitivity at the expense of generality. 

	\textit{Fermi} operates in survey mode most of the time, with slewing at $\approx$~4 degrees
per minute.  This creates a continuously changing background, which should be
accounted for, since ADWO would be optimal
without directional changes (as it uses the
correlation between the detectors and channels). One possibility would be to take the detailed satellite positional information into account and create 
a physical model to determine the background for a hundreds of seconds
\citep{2013A&A...557A...8S}. However, we expect that the slow slew will not suppress the sensitivity to the kind of short ($\sim$sec) transients that we are looking for. Therefore, a much simpler, 6th order polynomial
background fit was subtracted for each channel and detector, similar to the method of \citet{2016arXiv160203920C}. We chose the typical background  
window to be $\approx (-200,500)$ s around the search window, depending on the CTTE data availability: 
this window can contain the majority of GBM's GRB ligtcurves and covers approx. $1/7^{th}$ of \textit{Fermi's} orbit.

\section{GRB150522B}

\begin{figure}
\centering
\includegraphics[width=0.93\hsize]{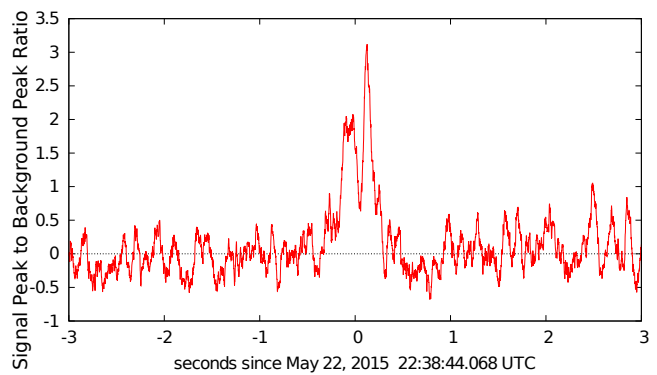}
\caption{ADWO light curve of GRB150522B in the 27-2000keV range.}
\label{fig:grb2}
\end{figure}

	To test ADWO, we analyze the short GRB150522B
gamma-ray burst, with T${}_{90}=1.02\pm~0.58$s	and
$2.13\pm~0.12\times10^{-7}$erg/cm${}^2$ fluence. These parameters are comparable
to the EM companion values of GW150914, as reported by \citet{2016arXiv160203920C}.  \textit{Fermi}
triggered on May 22, 2015 at 22:38:44.068 UTC, and full CTTE data of
$(-137,476)$s interval relative to the trigger is analyzed, using a 6~s long signal window centered on the trigger. The ADWO obtains a maximal SPBPR of 3.12,
and reveals the double pulse shown in the \textit{Fermi} GBM quicklook data (Fig.~\ref{fig:grb2}). 
The detector and energy channel weights are given in Tables~\ref{table:ei}-\ref{table:dj}. 

	To determine the significance we generated a Poisson-distributed synthetic
signal, using the background photon data of the interval, and repeated ADWO for
$10^4$ Monte-Carlo (MC) simulations with the same window width.  There was
no simulation with bigger SPBPR value than 3.12, therefore we estimate the false alarm rate 
to be below $2\times 10^{-5}~\mbox{Hz}$, and the false alarm probability to be below
$2\times 10^{-5}~\mbox{Hz} \times 0.125~\mbox{s} \times (1+\ln(6~\mbox{s}/64~\mbox{ms})) =2.8\times10^{-5}$, analogously to \citet{2016arXiv160203920C}.

\section{The GW150914 event}

\begin{figure}
\centering
\includegraphics[width=0.93\hsize]{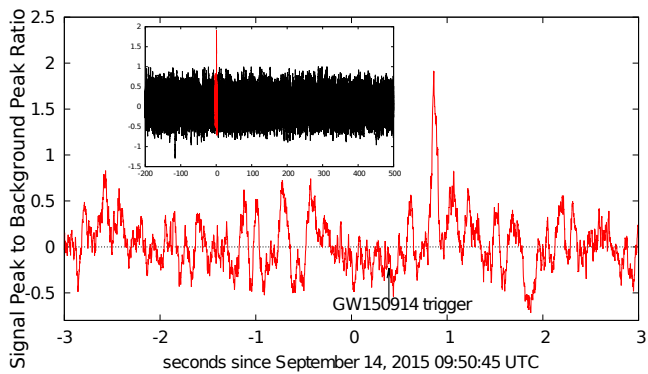}
\caption{ ADWO light curve of GW150914 in the 27-2000keV range. The inset shows the whole time interval where the ADWO search was performed.} 
 \label{fig:gw1}
\end{figure}

	We apply the ADWO method on the \textit{Fermi} CTTE data set covering the event of GW150914: the 6~s long signal window was centered
on September 14, 2015 09:50:45 UTC (391ms before trigger). Here we investigate a $(690-6)$~s time background interval
that adds up as $195$~s before and $495$~s after the time of the possible event.
The ADWO has converged (Fig.~\ref{fig:gw1}) and the obtained maximal SPBPR is 1.911, 474~ms after the GW trigger.

	 We repeated ADWO for $10^4$ MC simulations using this data: 
86 cases had bigger SPBPR than 1.911.  The false
alarm rate is $0.0014~\mbox{Hz}$, giving a false alarm probability of
$2.8\times 10^{-3}~\mbox{Hz} \times 0.474~\mbox{s} \times (1+\ln(6~\mbox{s}/64~\mbox{ms} )) =0.0075$, which is 
higher than 0.0022, the value given by \citet{2016arXiv160203920C}.

\begin{table}
	\caption[]{Channel weights}
	\label{table:ei}      
	\begin{tabular}{lllllll}
    \hline
	transient       &$e_3$   &   $e_4$    &   $e_5$  &  $e_6$   &    $e_7$    & $e_8$ \\
\hline
\small	GRB150522B &  \small 0.090 &  \small 0.297 &  \small 0.315 &  \small 0.188 & \small 0.000 & \small  0.110 \\
\small	GW150914 &  \small 0.203 & \small  0.050 &  \small 0.056 &  \small 0.559 &  \small 0.110 &  \small 0.022 \\
\small	LVT151012 & \small  0.260 & \small 0.212 & \small 0.010 & \small 0.113 & \small 0.000 & \small 0.406 \\
	    \hline
	\end{tabular}
\end{table}

\begin{table*}
	\caption{Detector weights for the $n0 \dots n9, na$ and $nb$ NaI(Tl) and $b0$ and $b1$ BGO detectors (as listed in \cite{2009ApJ...702..791M})}.
	\label{table:dj}      
	\centering          
	\begin{tabular}{lllllllllllllll}
	\hline       
	transient       &$n0$   &$n1$   &$n2$   &$n3$   &$n4$   &$n5$   &$n6$   &$n7$   &$n8$   &$n9$   &$na$   &$nb$   &$b0$   &$b1$   \\
    \hline
\small    GRB150522B & \small 0.105 & \small   0.106 & \small   0.100 & \small   0.078 & \small   0.146 & \small   0.073 & \small   0.001 & \small   0.031 & \small   0.000 & \small   0.021 & \small   0.009 & \small   0.050 & \small   0.113 & \small   0.167 \\
\small	GW150914 & \small 0.000& \small 0.044& \small 0.028& \small 0.151& \small 0.000& \small 0.000& \small 0.035& \small 0.045& \small 0.228& \small 0.090& \small 0.138& \small 0.162& \small 0.000& \small 0.077\\
\small	LVT151012 & \small  0.034 & \small   0.062 & \small   0.000 & \small   0.127 & \small   0.073 & \small   0.125 & \small   0.151 & \small   0.000 & \small   0.000 & \small   0.010 & \small   0.234 & \small   0.162 & \small   0.000 & \small   0.022 \\
		\hline
	\end{tabular}
\end{table*}

\section{LVT151012}

\begin{figure}
\centering
\includegraphics[width=0.93\hsize]{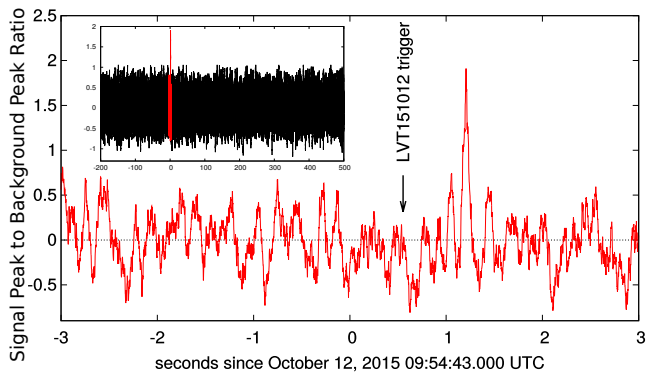}
\caption{ADWO light curve of LVT151012 in the 27-2000keV range. The inset shows the whole time interval where the ADWO search was performed. }
 \label{fig:lvt}
\end{figure}

Considering the positional errors of GW150914 on
the sky, it can be easily shown that there's a high ($>70-75\%$) probability
that a similar error ring will intersect with \textit{Fermi} GBM's field of view in the case of LVT151012 too. Therefore, we apply ADWO on the \textit{Fermi} GBM CTTE data around the event of LVT151012, covering $(-195,495)$~s, centered
on October 12, 2015 at 09:54:43 UTC.  We find a relatively strong signal at 09:54:44.207~UTC in the 6~s signal window, with a SPBPR of 1.805 (Fig.~\ref{fig:lvt}). 
The sum of the $27-290$~keV weights is higher than in the case of GW150914, i.e. here the signal is softer than GW150914 at the peak ($E_p\approx 3.5$~MeV), but harder than GRB15522 at the peak ($E_p\approx 130$~keV).  

	We made $10^4$ MC simulations: 308 cases had bigger SPBPR than the original observation. hence the false alarm rate is $0.0051~\mbox{Hz}$, 
and the false alarm probability is estimated to be 
$0.01~\mbox{Hz} \times 0.652~\mbox{s} \times (1+\ln(6~\mbox{s}/64~\mbox{ms} )) =0.037$.

When cross-checking the lightning detections made by WWLLN \citep{WWLLN} with the \textit{Fermi}'s positions and times, we find no TGF candidates (storm activity) within 500km of the spacecraft position and $\pm900~$s around the peak.

\section{Discussion}

Although here we applied our ADWO method to look for particular events, we point out that it is entirely possible to use this unsupervised data analysis method for a general search for non-triggered, short-duration \textit{Fermi} GBM events. 
Automatized search processes are important, as the total data-set collected by the \textit{Fermi}'s 8-years operation is significantly larger than the triggered data-set. 
It is likely that there are several potential EM events observed but not triggered. For example, based on the CTIME 256ms data product, \citet{2012grb..confE..36G} estimates $\approx 1.6$ untriggered SGRB/month in the \textit{Fermi} observations.
It is thus a worthwhile future task to identify potential SGRB candidates in the non-triggered \textit{Fermi} GBM data-set using ADWO. Alternatively, we can cross-check those already found by other algorithms.

As our ADWO method is independently developed, and only relies on the raw data of the satellite, it can provide a strong, independent test to any future signal. In regard of the current expectation that LIGO will detect several GW events in the near future, many of which may have a weak EM transient counterpart such as a SGRB, it is of crucial importance to identify those potential EM signals. We therefore expect that ADWO will be successfully applied in the future to find SGRB counterparts of the GW events observed by LIGO. 
The analysis of the GW151226 event as well as the improvement of ADWO is the topic of a forthcoming paper.

  
\begin{acknowledgements}
      This work was supported by OTKA grants NN111016 and NN114560. The authors wish to thank the World Wide Lightning Location Network (http://wwlln.net), a collaboration among over 50 universities and institutions, for providing the lightning location data used in this paper. Thanks for the computational resources of the Wigner GPU Laboratory of the Wigner RCP of the H.A.S,
A.C. Elbeze for pointing out that GRB150522B was the second GRB of the day
and the anonymous referee for valuable comments on the paper.

\end{acknowledgements}

\bibliographystyle{aa} 
\bibliography{semclgwfgd}

\end{document}